\documentclass{article}

\usepackage{spconf,amsmath,graphicx}
\usepackage{amssymb}
\usepackage{amsthm}
\usepackage{bm}
\usepackage{array}
\usepackage{float}
\usepackage{breqn}
\usepackage{multirow}
\usepackage{url}

\usepackage{bm}
\usepackage{amssymb,amsmath,graphicx,bbm,color,booktabs}
\usepackage{amsopn}

\newcommand{\va}{\bm{a}}

\newcommand{\vw}{\bm{w}}

\newcommand{\vphi}    {\bm{\phi}}

    \newcommand{\Lc}{\mathcal{L}}

  \newcommand{\Sc}{\mathcal{S}}

    \newcommand{\Yc}{\mathcal{Y}}

\newcommand{\R}{\mathbb{R}}

\newcommand{\reals}{\mathbb{R}}

\DeclareMathOperator*{\argmax}{argmax}

\newcommand{\sx}{\bar{\x}}

\newcommand{\sy}{\bar{\y}}
\newcommand{\spp}{\bar{\p}}
\newcommand{\syh}{\sy'}

\renewcommand{\eqref}[1]{Eq.~(\ref{#1})}

\def \x{{\mathbf x}}
\def \y{{\mathbf y}}

\def \p{{\mathbf p}}

\title{Phoneme Boundary Detection using Learnable Segmental Features}
\newcommand{\segfeat}{\textsc{SegFeat}~}
\newcommand{\bin}{\textsc{Bin}~}
\newcommand{\phn}{\textsc{Phn}~}

\name{Felix Kreuk$^1$, Yaniv Sheena$^2$~\thanks{Work was conducted while Yaniv was at Bar-Ilan University}, Joseph Keshet$^1$, Yossi Adi$^3$}
\address{$^1$Bar-Ilan University, Ramat-Gan, Israel \\ $^2$Facebook Inc. \\ $^3$Facebook AI Research }

\begin{document}

\maketitle

\begin{abstract}
Phoneme boundary detection plays an essential first step for a variety of speech processing applications such as speaker diarization, speech science, keyword spotting, etc. In this work, we propose a neural architecture coupled with a parameterized structured loss function to learn segmental representations for the task of phoneme boundary detection. First, we evaluated our model when the spoken phonemes were not given as input. Results on the TIMIT and Buckeye corpora suggest that the proposed model is superior to the baseline models and reaches state-of-the-art performance in terms of F1 and R-value. We further explore the use of phonetic transcription as additional supervision and show this yields minor improvements in performance but substantially better convergence rates. We additionally evaluate the model on a Hebrew corpus and demonstrate such phonetic supervision can be beneficial in a multi-lingual setting.

\end{abstract}
\begin{keywords}
Sequence segmentation,  recurrent neural networks (RNNs), structured prediction, phoneme boundary detection
\end{keywords}
\vspace{-0.2cm}
\section{Introduction}
\vspace{-0.2cm}
\emph{Phoneme Boundary Detection} or \emph{Phoneme Segmentation} plays an essential first step for a variety of speech processing applications such as speaker diarization~\cite{moattar2012review}, speech science~\cite{adi2016vowel,adi2015vowel}, keyword spotting~\cite{keshet2009discriminative}, Automatic Speech Recognition~\cite{kubala1996transcribing, rybach2009audio}, etc. 

Such segmentations are often modeled by either supervised or unsupervised methods. In the unsupervised setting, only the speech signal is given as input ~\cite{michel2016blind, dusan2006relation}, whereas in the supervised setting, the speech signal is accompanied by the target boundaries as supervision. Under the supervised setting, the set of pronounced or presumed phonemes is often provided additionally as input. This setting is referred to as \emph{forced alignment}. In cases where no phonemes are provided, we often denote the setup as \emph{text-independent} phoneme segmentation~\cite{aversano2001new, chen2016text}. In this study, we focus on the latter, namely phoneme boundary detection with no phonetic features. 

Inspired by the work on learning segmental features~\cite{kiperwasser2016simple, adi2017sequence, sha2005real}, we suggest learning segmental representations using neural models for both phoneme boundaries and phoneme segments to detect phoneme boundaries accurately.

Specifically, we jointly optimize a Recurrent Neural Network (RNN) and structured loss parameters by using RNN outputs as feature functions for a structured model. First, the RNN encodes the entire speech utterance and outputs a new representation for each frame. Then, an efficient search is applied over all possible segments so that the most probable one can be selected. To compare with prior work, we evaluate this approach using TIMIT~\cite{garofolo1993timit} and Buckeye~\cite{pitt2005buckeye} datasets. The proposed approach outperforms the baseline models on both datasets and reaches State-of-the-Art (SOTA) performance. We additionally experiment with leveraging phoneme information as secondary supervision to the model and demonstrate that such an approach can be beneficial for both performance and convergence speed. Lastly, we demonstrated that such supervision could be beneficial when detecting phoneme boundaries from different languages, by training on English and testing on Hebrew.

\vspace{-0.2cm}
\section{Related Work}
\vspace{-0.2cm}

The problem of phoneme boundary detection was explored in various settings. In the unsupervised setting, also known as \emph{blind phoneme segmentation}, the speech signal is provided by itself with no additional phonemes nor boundaries as supervision. Traditionally, signal processing techniques were used to find spectral changes in the signal to detect phoneme boundaries~\cite{dusan2006relation,estevan2007finding,hoang2015blind,almpanidis2008phonemic, rasanen2011blind} and the references therein. Recently, the authors of~\cite{michel2016blind} suggested training a next-frame prediction model using either an approximated Markov model or RNN to identify potential phoneme transitions at regions of high error.

Under the supervised setting, the most common approach is the forced alignment setup. Models that follow such an approach involve with Hidden Markov Models or different structured prediction algorithms using handcrafted features~\cite{keshet2005phoneme, mcauliffe2017montreal}. The main drawback of forced alignment methods is the need for phoneme annotations also at inference time. This requirement may limit the applicability of such techniques to a monolingual setting due to mismatched phoneme sets, i.e., it is unclear how to handle unseen phonemes from a foreign language.

Another supervised setting is the text-independent phoneme boundary detection, in which the model is provided with phoneme boundaries as supervision but without information about the uttered phonemes. Most previous works that follow such setup consider the problem as a binary classification task where one label is associated with phoneme boundaries annotations, and another one for the rest of the signal. The authors in~\cite{king2013accurate} used a kernel-based method, composed of six radial basis function support vector machines under such setup, while the authors in~\cite{franke2016phoneme} suggested using RNN followed by a peak detection algorithm to predict phoneme boundaries. The main drawback of such modeling is that it ignores the internal structure of the output while considering boundaries to be conditionally independent.

Another related lines of work to be noticed is the work of learnable segmental features in ASR~\cite{lu2016segmental}, dependency parsing~\cite{kiperwasser2016simple}, and word segmentation~\cite{adi2017sequence}.

\vspace{-0.2cm}
\section{Problem Setting}
\vspace{-0.2cm}

In the problem of phoneme boundary detection we are provided with a speech utterance, denoted as $\sx = (\x_1,\ldots,\x_T)$, represented as a sequence of acoustic feature vectors, where each $\x_t\in\R^D$ $(1\leq t \leq T)$ is a $D$-dimensional vector. The length of the speech utterance, $T$, is not a fixed value, since input utterances can have different duration. 

Each input utterance is associated with a timing sequence, denoted by $\sy = (y_1,\ldots,y_k)$, where $k \leq T$ is the number of segments, and can vary across different inputs. Each element $y_i \in \Yc$, where $\Yc=\{1,\ldots,T\}$  indicates the start time of a new event in the speech signal. We denote by $\Yc^*$ the set of all finite-length sequences over $\Yc$. Although $\sy \in \Yc^*$, can be of any length, we assume its length to be bounded by $T$.

\vspace{-0.3cm}
\section{Model Description}
\vspace{-0.3cm}
Following the structured prediction framework~\cite{adi2016structed}, also known as Energy Based Models~\cite{lecun2006tutorial}, consider the following prediction rule with $\vw \in \reals^d$, such that $\syh_{\vw}$ is a good approximation to the true label of $\sx$, as follows:
\begin{dmath}
\label{eq:yw}
\syh_{\vw}(\sx) = \argmax_{\sy \in \Yc^*} ~ \vw^\top \vphi(\sx, \sy)
\end{dmath} 
Where $\vphi$ is a mapping function from the set of input objects and target labels to a real vector of length $d$. 
We assume there exists some unknown probability distribution $\rho$ over pairs $(\sx, \sy)$ where $\sy$ is the desired output (or reference output) for $\sx$. We are provided with a training set $\Sc = \{(\sx_1,\sy_1),\ldots,(\sx_m,\sy_m)\}$ of $m$ examples that are drawn i.i.d. from $\rho$. Our goal is to set $\vw$ so as to minimize the expected empirical cost, or the \emph{empirical risk},
\begin{dmath}
\label{eq:reg-loss}
\Lc(\vw, \sx, \sy) = \frac{1}{m}\sum_{i=1}^{m} \ell(\vw, \sx, \sy)
\end{dmath} 
where we define $\ell(\vw, \sx, \sy)$ to be the hinge loss function as follows,
\[ 
\ell(\vw, \sx, \sy) = \max_{\syh \in \Yc^*} ~ \left[1 - \vw^\top\vphi(\sx, \sy) + \vw^\top\vphi(\sx,\syh_{\vw}) \right]
\]

We assume $\vphi(\sx, \sy)$ can be decomposed as $\sum_{i=1}^{k} \vphi'(\sx, y_i)$ where each $\vphi'$ can be extracted using different techniques, e.g., hand-crafted, feed-forward neural network, RNNs, etc. Traditionally, $\vphi'$ was manually chosen using data analysis techniques and involved manipulations of local and global features.

Notice, such decomposition implicitly assumes the output boundaries to be conditionally independent. However, considering the previous boundary prediction provides a lot of information regarding the next boundary. For example, when predicting a vowel offset boundary, considering the vowel onset boundary may provide some insight regarding the typical vowel length. An example of such segmental features might be the mean energy of the spectrum or the average fundamental frequency, $f_0$, value in specific ranges~\cite{adi2016vowel}. Hence, we adjust the objective as follows,

\begin{equation}
\begin{split}
\label{eq:dec_phi}
\syh_{\vw}(\sx) 
& = \argmax_{\sy \in \Yc^*} ~ \vw^\top \vphi(\sx, \sy) \\
& = \argmax_{\sy \in \Yc^*} ~ \vw^\top \Big( \sum_{i=1}^{k} \vphi'_u(\sx, y_i) + \sum_{j=1}^{k-1} \vphi'_{bi}(\sx, y_j, y_{j+1})\Big) \\
\end{split}
\end{equation} 
\vspace{-0.1cm}

\begin{figure}[t]
\centering   
\includegraphics[scale=0.4]{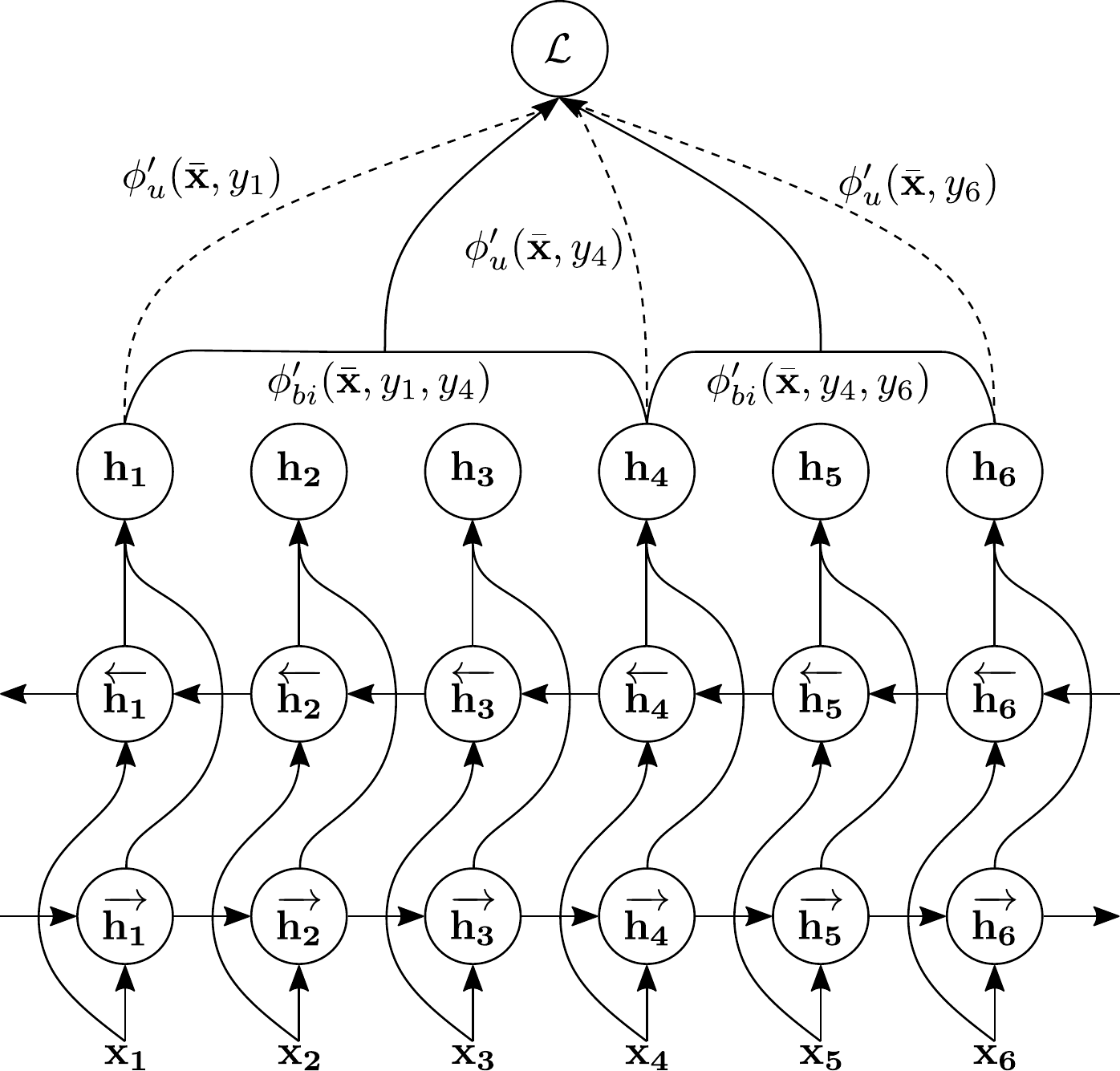}
\caption{An illustration for using BI-RNN as feature functions. We search through all possible locations and predict the one with the highest score. In this example the score is calculated for timing sequence is $(1, 4, 6)$.}
 \label{fig:rnn}
 \vspace{-0.4cm}
\end{figure}

In this work, we consider $\vphi'_u$ and $\vphi'_{bi}$ as \emph{Learnable Segmental Features}. One can explore various types of neural network architectures to model $\vphi'_u$ and $\vphi'_{bi}$. We set $\vphi'_u(\sx, y_i)$ to be a bidirectional RNN output at $y_i$, and $\vphi'_{bi}(\sx, y_j, y_{j+1})$ as the sum of the RNN outputs between $y_j$ and $y_{j+1}$. An illustration of the proposed model depicted in Figure~\ref{fig:rnn}. Notice that the $\argmax$ operator in Equation \ref{eq:dec_phi} may search over exponential number of target segmentations. Luckily, this can be implemented efficiently using a dynamic-programming algorithm, and refer the reader to our implementation for more details. The source code will be available under the following repository: \url{hyperurl.co/2w206g}

Our goal is to find the model parameters so as to minimize the risk as in Equation \ref{eq:reg-loss}. Since all components described above are differentiable we can jointly optimize them using gradient based methods such as stochastic gradient descent (SGD). We denote the proposed model as $\segfeat$.

\vspace{-0.2cm}
\section{Experiments}
\vspace{-0.2cm}

\begin{table}[t]
	\renewcommand{\arraystretch}{1.45}
	\small
	\centering
	\caption{Comparison of phoneme segmentation models. Precision (P) and recall (R) are calculated with tolerance value of 20 ms}
  \vspace{5pt}
  \label{tab:phoneme_seg_results}
  \begin{tabular}{l|l|cccc}
	\toprule
	& Model & P & R & F1 & R-val  \\
    \midrule
    \multirow{3}{*}{ \rotatebox[origin=c]{90}{TIMIT} }
	  & King \emph{et al.}\cite{king2013accurate}  &  87.0 & 84.8 & 85.9 & 87.8   \\
      & Franke \emph{et al.}\cite{franke2016phoneme}  & 91.1 & 88.1 &  89.6 & 90.8  \\
      & \segfeat   & \textbf{94.03} & \textbf{90.46}  & \textbf{92.22}   & \textbf{92.79} \\
	\midrule
      \multirow{2}{*}{ \rotatebox[origin=c]{90}{Buckeye} }
      & Franke \emph{et al.}\cite{franke2016phoneme}  & \textbf{87.8} & 83.3 &  85.5 & 87.17  \\
      & \segfeat   & 85.4 & \textbf{89.12}  & \textbf{87.23}  & \textbf{88.76} \\
    \bottomrule
  \end{tabular}
\end{table}

In this section, we provide a detailed description of the experiments. We start by presenting the experimental setting. Then we outline the evaluation method. We conclude this section with the boundary detection results and ablation.

\vspace{-0.2cm}
\subsection{Experimental setup}
\vspace{-0.2cm}

We implemented our model using a 2-layer bidirectional LSTM network. The final score outputs $\vphi'_{u}$ and $\vphi'_{bi}$ were calculated using a 2-layer fully-connected network that follows the RNN. The proposed model was trained for 150 epochs using Adam optimizer with a learning rate of $1e^{-4}$.

We evaluated our model on both TIMIT and Buckeye corpora. For the TIMIT corpus, we used the standard train/test split, where we randomly sampled ~10\% of the training set for validation. For Buckeye, we split the corpus at the speaker level into training, validation, and test sets with a ratio of 80/10/10. Similarly to~\cite{franke2016phoneme}, we split long sequences into smaller ones by cutting during noises, silences, and un-transcribed segments. Overall, each sequence started and ended with a maximum of 20 ms of non-speech\footnote{All experiments using Buckeye dataset were done at Bar-Ilan university.}. For both corpora, we extracted 13 Mel-Frequency Cepstrum Coefficients (MFCCs), with delta and delta-delta features every 10 ms, with a processing window size of 10ms. 

Moreover, we concatenated four additional features based on the spectral changes between adjacent frames, using MFCCs to represent the spectral properties of the frames. Define  $D_{t,j}=d(\va_{t-j},\va_{t+j})$ to be the Euclidean distance between the MFCC feature vectors $\va_{t-j}$ and $\va_{t+j}$, where $\va_t\in\R^{39}$ for $1 \le t \le T$. The features are denoted by $D_{t,j}$, for $j\in \{1, 2, 3, 4\}$. We observed this set of features greatly improves performance over the standard MFCC features.

\vspace{-0.2cm}
\subsection{Evaluation method}
\vspace{-0.2cm}
Following previous work on phoneme boundary detection~\cite{dusan2006relation, franke2016phoneme, king2013accurate, michel2016blind}, we evaluated the performance of the proposed models and baseline models using precision ($P$), recall ($R$) and F1-score with a tolerance level of 20 ms. A drawback of the F1-score measurement is that a high recall and a low precision might yield relatively a high F1-score (e.g., see Section 4.2 in~\cite{michel2016blind}). The authors in~\cite{rasanen2009improved} proposed a complementary metric denoted as \emph{R-value}:
\begin{equation}
\begin{split}
&\textrm{R-value} = 1 - \frac{|r_1|+|r_2|}{2} \\
&r_1 = \sqrt{(1-R)^2 + (OS)^2}, ~~  
r_2 = \frac{(-OS+R-1)}{\sqrt{2}}
\end{split}
\end{equation}
where $OS$ is an over-segmentation measure, defined as $OS=R/P-1$. Overall the performance is presented in terms of Precision, Recall, F-score and R-value.

\vspace{-0.2cm}
\subsection{Results}
\vspace{-0.2cm}
In Table~\ref{tab:phoneme_seg_results} we compare the proposed model against two baselines: Franke \emph{et al.}~\cite{franke2016phoneme}, and King \emph{et al.}~\cite{king2013accurate}. Results suggest that our proposed model is superior to the baseline models over all metrics on both corpora with one exception of the Precision metric on the Buckeye corpus.

\begin{figure}[t]
\centering
\includegraphics[width=0.45\textwidth]{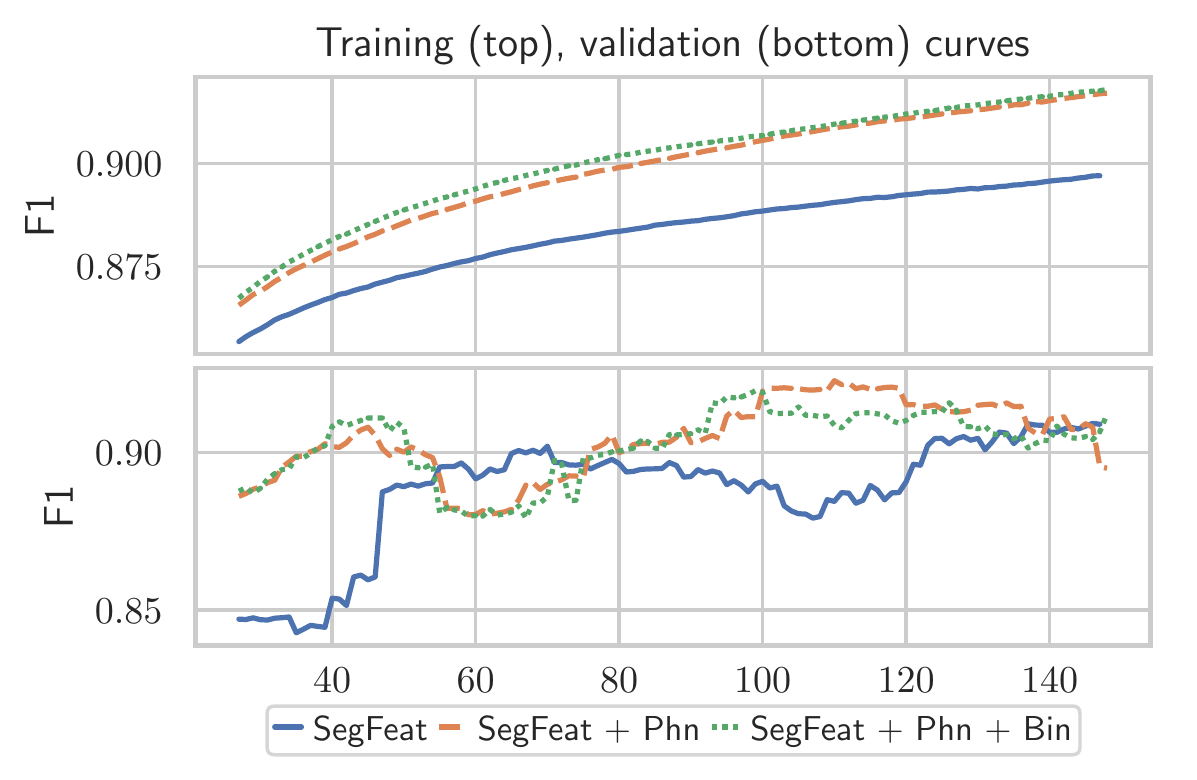}
\vspace{-0.3cm}
\caption{F1-score as a function of update steps for different models on the TIMIT validation set.}
\label{fig:train}
\end{figure}

In the next subsection, we conduct an ablation study to explore the effect of using phonetic annotations as additional supervision.

\vspace{-0.2cm}
\subsection{Ablation study}
\vspace{-0.2cm}
Recall that in the task of phoneme boundary detection, we are often provided with the pronounced unaligned phoneme sequence. For example, for the timing sequence $\sy=(19, 23, 40, 45)$, we additionally get the phoneme sequence $\spp=(\textrm{/sil/, /b/, /ae/, /d/, /sil/})$. As opposed to the forced alignment setup where phonemes are used as additional inputs to the model, one could leverage the provided phonemes as supervision. Such annotations should be provided during the training phase only; hence, they do not restrict the model by any phonetic features. To explore the effect of using phoneme annotations as another supervision we augment the original training objective in Eq.~\ref{eq:reg-loss} with a phoneme classification loss (Negative Log-likelihood) at each time frame, and denote this loss function by \phn. For completeness, we additionally conducted experiments with including a binary boundary detection loss as suggested in~\cite{franke2016phoneme, king2013accurate}, and denote it by \bin.

We trained several models while optimizing different sets of loss functions on the TIMIT corpus. Results are summarized in Table~\ref{tab:ablation}. Such comparison sheds light on the effect each loss function has on model performance. 
 
\begin{table}[t]
    \small
	\centering
	\vspace{-0.3cm}
	\caption{Models performance on TIMIT using different sets of loss function.}
  \vspace{5pt}
  \label{tab:ablation}
  \begin{tabular}{l|cccc}
	\toprule
	Loss & P & R & F1 & R-val   \\
    \midrule
       \bin & 91.1 & 88.1 &  89.6 & 90.8  \\
       \bin + \phn  & \textbf{96.6} & 85.0 &  90.04 & 89.33  \\
       \segfeat                                    & 94.03 & 90.46 & 92.22 & 92.79 \\
       \segfeat +\phn                              & 92.98 & 92.33 & 92.66 & 93.69 \\
       \segfeat +\phn +\bin                        & 92.67 & \textbf{93.03} & \textbf{92.85} & \textbf{93.91} \\
    \bottomrule
  \end{tabular}
  \vspace{-0.3cm}
\end{table}
\vspace{-0.2cm}
\subsection{Forced alignment comparison}
\vspace{-0.2cm}
Lastly, in this subsection we compare our model against SOTA forced-alignment models \cite{keshet2005phoneme, mcauliffe2017montreal}. Recall, forced-alignment is defined as the task of finding the phoneme boundaries of a given set of phonemes, this task may be supervised or unsupervised. Notice that while these models are provided with the pronounced phoneme sequence at test time, the only input to our model is the speech signal. All models were trained and tested on TIMIT dataset. Results are summarized in Table \ref{tab:forced}.
\begin{table}[t]
	\vspace{-0.4cm}
    \small
	\centering
	\caption{Comparison of the proposed model against forced-alignment algorithms.}
  \vspace{5pt}
  \label{tab:forced}
  \begin{tabular}{l|ccccc}
    \toprule
	Model & P & R & F1 & R-val  \\
    \midrule
McAuliffe~(unsup.)~\cite{mcauliffe2017montreal}   & 83.9 & 81.6 & 82.7 & 85.16 \\
Keshet~(sup.)~\cite{keshet2005phoneme}   & 90 & 82.2 & 85.9 & 79.51 \\    
       \segfeat & \textbf{94.03} & \textbf{90.46}  & \textbf{92.22}   & \textbf{92.79} \\
    \bottomrule
  \end{tabular}
\end{table}

Leveraging phoneme supervision under the binary boundary detection setting, i.e., optimizing both \phn and \bin loss functions, reaches the best Precision score. This, however, comes at the cost of the worst Recall and R-value. Contrarily, leveraging such supervision using learnable segmental features yields the best Recall, F1-score, and R-value. 

While all $\segfeat$ models reached a comparable F1-score on the test set, incorporating phoneme supervision improves R-value and convergence speed by $\sim$30\%. Additionally, adding the \bin loss function to the optimization had a minor effect on the model performance. Figure \ref{fig:train} depicts F1-score as a function of model iterations for different models. 

\begin{table}[t]
\vspace{-0.4cm}
    \small
	\centering
	\caption{An ablation study on the effect of the \phn loss on Hebrew language.}
  \vspace{5pt}
  \label{tab:heb}
  \begin{tabular}{l|ccccc}
    \toprule
	Model & P & R & F1 & R-val  \\
    \midrule
       \segfeat w/o~  \phn Loss   & \textbf{83.58} & 79.2 & 81.24 & 83.67 \\
       \segfeat w~ \phn Loss   & 83.11 & \textbf{81.66} & \textbf{82.38} & \textbf{84.92} \\
    \bottomrule
  \end{tabular}
  \vspace{-0.1cm}
\end{table}

\begin{figure}[t]
\centering
\includegraphics[width=0.45\textwidth]{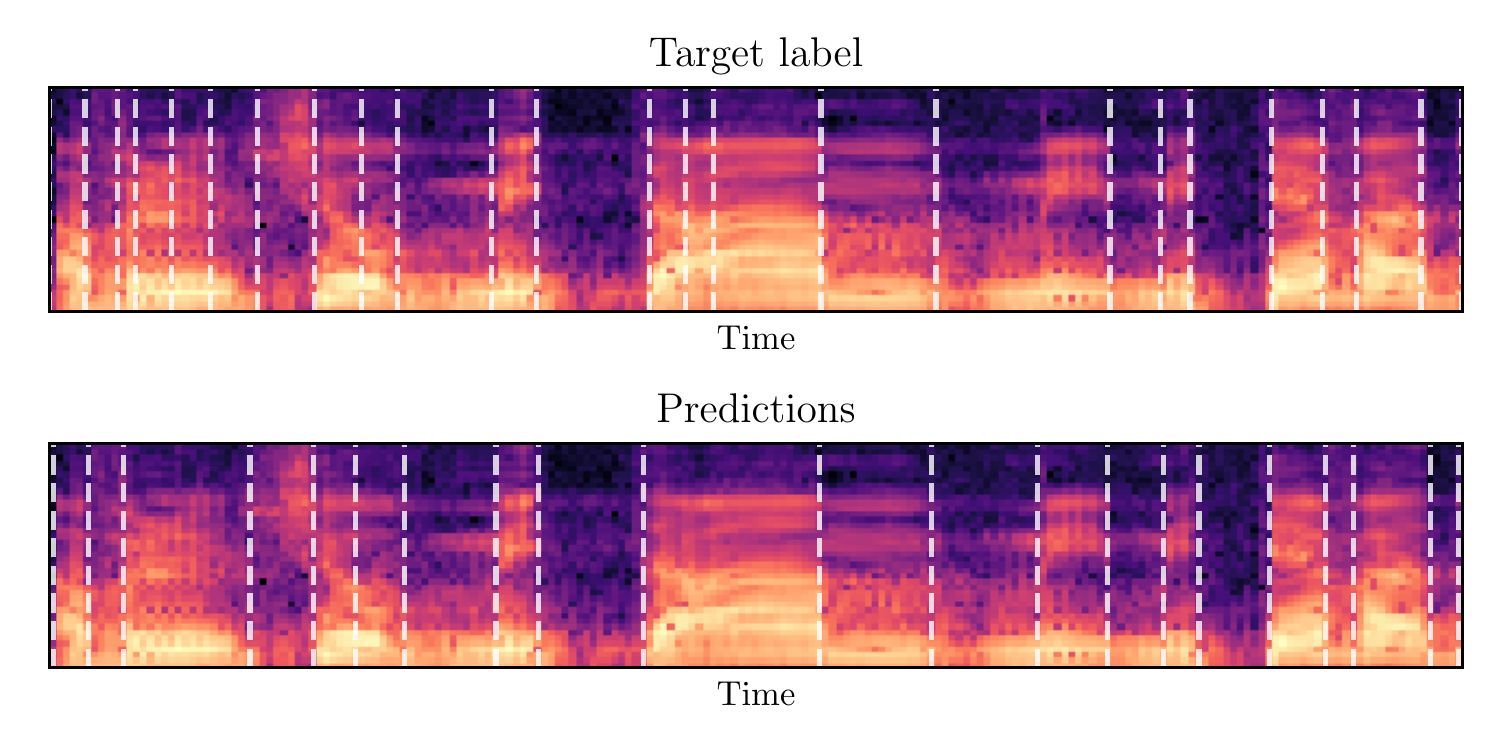}
\vspace{-0.5cm}
\caption{Example of segmentation result on an Hebrew utterance using an English trained model.}
\label{fig:spect}
\end{figure}

\vspace{-0.2cm}
\subsection{Cross-language phoneme boundary detection}
\vspace{-0.2cm}

One concern when using the phoneme classification loss is convergence to a language-dependent model. To further explore that we evaluated the performance of two models that were trained on American English (TIMIT) and tested on a Hebrew language corpus~\cite{benautomatic}. The Hebrew language has only partial overlap in its phonetic set to English, and the utterances in the Hebrew corpus consist of both read and spontaneous speech. Results are summarized in Table \ref{tab:heb}.

\noindent Interestingly, results suggest that incorporating the phoneme classification loss with segmental features slightly improved overall performance in comparison to using segmental features alone. An example of the model predictions together with the target label is depicted in Figure~\ref{fig:spect}.


\vspace{-0.3cm}
\section{Discussion and Future work}
\vspace{-0.3cm}

We suggest using learnable segmental features for the task of phoneme boundary detection. We showed that such a phoneme-independent approach is superior to the baseline methods and can reach SOTA results on the TIMIT and Buckeye corpora. Moreover, we demonstrated that such a text-independent model could be efficiently applied to the Hebrew language, while it was trained on American English data.

For future work, we would like to explore the cross-domain setting further and while providing a systematic comparison between different languages. Additionally, we would like to examine the usability of the model for low-resource settings.

\bibliographystyle{IEEEbib}
\bibliography{mybib}

\begin{thebibliography}{10}

\bibitem{moattar2012review}
Mohammad~H Moattar and Mohammad~M Homayounpour,
\newblock ``A review on speaker diarization systems and approaches,''
\newblock {\em Speech Communication}, vol. 54, no. 10, pp. 1065--1103, 2012.

\bibitem{adi2016vowel}
Yossi Adi, Joseph Keshet, Emily Cibelli, Erin Gustafson, Cynthia Clopper, and
  Matthew Goldrick,
\newblock ``Automatic measurement of vowel duration via structured
  prediction,''
\newblock {\em The Journal of the Acoustical Society of America}, vol. 140, no.
  6, pp. 4517--4527, 2016.

\bibitem{adi2015vowel}
Yossi Adi, Joseph Keshet, and Matthew Goldrick,
\newblock ``Vowel duration measurement using deep neural networks,''
\newblock in {\em 2015 IEEE 25th International Workshop on Machine Learning for
  Signal Processing (MLSP)}. IEEE, 2015, pp. 1--6.

\bibitem{keshet2009discriminative}
Joseph Keshet, David Grangier, and Samy Bengio,
\newblock ``Discriminative keyword spotting,''
\newblock {\em Speech Communication}, vol. 51, no. 4, pp. 317--329, 2009.

\bibitem{kubala1996transcribing}
Francis Kubala, Tasos Anastasakos, Hubert Jin, Long Nguyen, and Richard
  Schwartz,
\newblock ``Transcribing radio news,''
\newblock in {\em Proceeding of Fourth International Conference on Spoken
  Language Processing. ICSLP'96}. IEEE, 1996, vol.~2, pp. 598--601.

\bibitem{rybach2009audio}
David Rybach, Christian Gollan, Ralf Schluter, and Hermann Ney,
\newblock ``Audio segmentation for speech recognition using segment features,''
\newblock in {\em 2009 IEEE International Conference on Acoustics, Speech and
  Signal Processing}. IEEE, 2009, pp. 4197--4200.

\bibitem{michel2016blind}
Paul Michel, Okko R{\"a}s{\"a}nen, Roland Thiolliere, and Emmanuel Dupoux,
\newblock ``Blind phoneme segmentation with temporal prediction errors,''
\newblock {\em arXiv preprint arXiv:1608.00508}, 2016.

\bibitem{dusan2006relation}
Sorin Dusan and Lawrence Rabiner,
\newblock ``On the relation between maximum spectral transition positions and
  phone boundaries,''
\newblock in {\em Ninth International Conference on Spoken Language
  Processing}, 2006.

\bibitem{aversano2001new}
Guido Aversano, Anna Esposito, and M~Marinaro,
\newblock ``A new text-independent method for phoneme segmentation,''
\newblock in {\em Proceedings of the 44th IEEE 2001 Midwest Symposium on
  Circuits and Systems. MWSCAS 2001 (Cat. No. 01CH37257)}. IEEE, 2001, vol.~2,
  pp. 516--519.

\bibitem{chen2016text}
Lijiang Chen, Xia Mao, and Hong Yan,
\newblock ``Text-independent phoneme segmentation combining egg and speech
  data,''
\newblock {\em IEEE/ACM transactions on audio, speech, and language
  processing}, vol. 24, no. 6, pp. 1029--1037, 2016.

\bibitem{kiperwasser2016simple}
Eliyahu Kiperwasser and Yoav Goldberg,
\newblock ``Simple and accurate dependency parsing using bidirectional lstm
  feature representations,''
\newblock {\em Transactions of the Association for Computational Linguistics},
  vol. 4, pp. 313--327, 2016.

\bibitem{adi2017sequence}
Yossi Adi, Joseph Keshet, Emily Cibelli, and Matthew Goldrick,
\newblock ``Sequence segmentation using joint rnn and structured prediction
  models,''
\newblock in {\em 2017 IEEE International Conference on Acoustics, Speech and
  Signal Processing (ICASSP)}. IEEE, 2017, pp. 2422--2426.

\bibitem{sha2005real}
Fei Sha and Lawrence~K Saul,
\newblock ``Real-time pitch determination of one or more voices by nonnegative
  matrix factorization,''
\newblock in {\em Advances in Neural Information Processing Systems}, 2005, pp.
  1233--1240.

\bibitem{garofolo1993timit}
J.~S Garofolo,
\newblock ``Timit acoustic phonetic continuous speech corpus,''
\newblock {\em Linguistic Data Consortium, 1993}, 1993.

\bibitem{pitt2005buckeye}
M.~A Pitt et~al.,
\newblock ``The buckeye corpus of conversational speech: Labeling conventions
  and a test of transcriber reliability,''
\newblock {\em Speech Communication}, vol. 45, no. 1, pp. 89--95, 2005.

\bibitem{estevan2007finding}
Yago~Pereiro Estevan, Vincent Wan, and Odette Scharenborg,
\newblock ``Finding maximum margin segments in speech,''
\newblock in {\em 2007 IEEE International Conference on Acoustics, Speech and
  Signal Processing-ICASSP'07}. IEEE, 2007, vol.~4, pp. IV--937.

\bibitem{hoang2015blind}
Dac-Thang Hoang and Hsiao-Chuan Wang,
\newblock ``Blind phone segmentation based on spectral change detection using
  legendre polynomial approximation,''
\newblock {\em The Journal of the Acoustical Society of America}, vol. 137, no.
  2, pp. 797--805, 2015.

\bibitem{almpanidis2008phonemic}
George Almpanidis and Constantine Kotropoulos,
\newblock ``Phonemic segmentation using the generalised gamma distribution and
  small sample bayesian information criterion,''
\newblock {\em Speech Communication}, vol. 50, no. 1, pp. 38--55, 2008.

\bibitem{rasanen2011blind}
Okko R{\"a}s{\"a}nen, Unto~K Laine, and Toomas Altosaar,
\newblock ``Blind segmentation of speech using non-linear filtering methods,''
\newblock {\em Speech Technologies}, pp. 105--124, 2011.

\bibitem{keshet2005phoneme}
Joseph Keshet, Shai Shalev-Shwartz, Yoram Singer, and Dan Chazan,
\newblock ``Phoneme alignment based on discriminative learning,''
\newblock in {\em Ninth European Conference on Speech Communication and
  Technology}, 2005.

\bibitem{mcauliffe2017montreal}
Michael McAuliffe, Michaela Socolof, Sarah Mihuc, Michael Wagner, and Morgan
  Sonderegger,
\newblock ``Montreal forced aligner: Trainable text-speech alignment using
  kaldi.,''
\newblock in {\em Interspeech}, 2017, pp. 498--502.

\bibitem{king2013accurate}
Sarah King and Mark Hasegawa-Johnson,
\newblock ``Accurate speech segmentation by mimicking human auditory
  processing,''
\newblock in {\em 2013 IEEE International Conference on Acoustics, Speech and
  Signal Processing}. IEEE, 2013, pp. 8096--8100.

\bibitem{franke2016phoneme}
Joerg Franke, Markus Mueller, Fatima Hamlaoui, Sebastian Stueker, and Alex
  Waibel,
\newblock ``Phoneme boundary detection using deep bidirectional lstms,''
\newblock in {\em Speech Communication; 12. ITG Symposium}. VDE, 2016, pp.
  1--5.

\bibitem{lu2016segmental}
Liang Lu, Lingpeng Kong, Chris Dyer, Noah~A Smith, and Steve Renals,
\newblock ``Segmental recurrent neural networks for end-to-end speech
  recognition,''
\newblock {\em arXiv preprint arXiv:1603.00223}, 2016.

\bibitem{adi2016structed}
Yossi Adi and Joseph Keshet,
\newblock ``Structed: risk minimization in structured prediction,''
\newblock {\em The Journal of Machine Learning Research}, vol. 17, no. 1, pp.
  2282--2286, 2016.

\bibitem{lecun2006tutorial}
Yann LeCun, Sumit Chopra, Raia Hadsell, M~Ranzato, and F~Huang,
\newblock ``A tutorial on energy-based learning,''
\newblock {\em Predicting structured data}, vol. 1, no. 0, 2006.

\bibitem{rasanen2009improved}
O.~J. R{\"a}s{\"a}nen, U.~K. Laine, and T.~Altosaar,
\newblock ``An improved speech segmentation quality measure: the r-value,''
\newblock in {\em Tenth Annual Conference of the International Speech
  Communication Association}, 2009.

\bibitem{benautomatic}
A.~Ben-Shalom, J.~Keshet, D.~Modan, and A.~Laufer,
\newblock ``Automatic tools for analyzing spoken hebrew,''
\newblock in {\em Afeka Conference for Speech Processing}, 2014.

\end{thebibliography}

\end{document}